\documentclass[a4paper,aps,prd,twocolumn,showpacs,preprintnumbers,nofootinbib,amsmath,amssymb, superscriptaddress]{revtex4-1}

\usepackage{array}
\usepackage{graphicx}
\usepackage{bm}
\usepackage{amsmath, amsthm,amssymb,url}
\usepackage[normalem]{ulem}
\usepackage{color}
\usepackage{subcaption}

\def\Msun{M_{\odot}}

\DeclareMathOperator*{\argmax}{arg\,max}

\def\be{\begin{equation}}
\def\ee{\end{equation}}
\def\bea{\begin{eqnarray}}
\def\eea{\end{eqnarray}}
\def\bn{\begin{enumerate}}
\def\en{\end{enumerate}}

\usepackage{color}
\makeatletter

\newcommand{\Rmnum}[1]{\expandafter\@slowromancap\romannumeral #1@}
\makeatother

\begin{document}
\captionsetup[figure]{labelfont={bf},labelformat={default},labelsep=period,name={Figure.}}
\preprint{------}

\title{Driving unmodelled gravitational-wave transient searches using astrophysical information} 

\author{P.~Bacon}
\affiliation{APC, Univ Paris Diderot, CNRS/IN2P3, CEA/Irfu, Obs. de Paris, Sorbonne Paris Cit\'{e}, F-75013 Paris, France}

\author{Gayathri~V.}
\affiliation{Department of Physics, Indian Institute of Technology Bombay, Powai, Mumbai, Maharashtra 400076, India}

\author{E.~Chassande-Mottin}
\affiliation{APC, Univ Paris Diderot, CNRS/IN2P3, CEA/Irfu, Obs. de Paris, Sorbonne Paris Cit\'{e}, F-75013 Paris, France}

\author{A.~Pai}
\affiliation{Department of Physics, Indian Institute of Technology Bombay, Powai, Mumbai, Maharashtra 400076, India}

\author{F.~Salemi}
\affiliation{Max Planck Institute for Gravitational Physics (Albert Einstein Institute), D-30167 Hannover, Germany}

\author{G.~Vedovato}
\affiliation{INFN, Sezione di Padova, I-35131 Padova, Italy}

\date{\today}
\begin{abstract}
Transient gravitational-wave searches can be divided into two main
families of approaches: modelled and unmodelled searches, based on
matched filtering techniques and time-frequency excess power
identification respectively. The former, mostly applied in the context
of compact binary searches, relies on the precise knowledge of the
expected gravitational-wave phase evolution. This information is not
always available at the required accuracy for all plausible
astrophysical scenarios, e.g., in presence of orbital precession, or
eccentricity.  The other search approach imposes little priors on the
targetted signal. We propose an intermediate route based on a
modification of unmodelled search methods in which time-frequency
pattern matching is constrained by astrophysical waveform models (but
not requiring accurate prediction for the waveform phase evolution).
The set of astrophysically motivated patterns is conveniently
encapsulated in a graph, that encodes the time-frequency pixels and
their co-occurrence. This allows the use of efficient graph-based
optimization techniques to perform the pattern search in the data.  We
show in the example of black-hole binary searches that such an
approach leads to an averaged increase in the distance reach (+7-8\%)
for this specific source over standard unmodelled searches.
\end{abstract}


\maketitle

\section{Introduction}
\label{SecI}

The first gravitational wave detections made by the advanced LIGO
detectors (namely GW150914 \cite{abbott2016:gw150914},
GW151226 \cite{abbott2016:gw151226},
GW170104 \cite{abbot17:_2017PhRvL.118v1101A} and GW170608 \cite{abbott17:_gw170}), and recently jointly with
advanced Virgo (GW170814 \cite{abbott2017:gw170814},
GW170817 \cite{abbott2017:gw170817}) heralds a new astronomy which
will develop further as more detectors come online such as
Kagra \cite{aso13:_inter_kagra} and LIGO
India \cite{iyer11:_ligo_india_propos_consor_indian}. So far, the
detections of the gravitational wave signals associated with the
merger of four binary black holes
(BBH) \cite{abbott16:_binar_black_hole_merger_first,abbot17:_2017PhRvL.118v1101A}
and a binary neutron star \cite{TheLIGOScientific:2017qsaGW170817} have been announced.

The search for BBH signals in the LIGO/Virgo data is performed using a
variety of methods, including the matched filtering technique 
(such as \textsc{PyCBC} \cite{pycbc2016}, \textsc{gstLAL} \cite{gstlal2014}
or \textsc{MBTA} \cite{Adams:2015ulm}). This pattern matching algorithm essentially consists in comparing 
the data with a waveform model or
``template''. The matched filtering BBH searches use a discrete grid
of templates. This grid samples the physical parameter space with
sufficient density to detect at least 90 \% of the detectable binary sources
(assuming the real signals do not deviate from the waveform model).  The
template grid used for the analysis of the first Advanced LIGO run
includes $\sim 250,000$ waveforms spanning the compact binaries with
total mass $M = m_1 + m_2 < 100 \Msun$, mass ratio $q = m_1/m_2 <
99$ and the dimensionless spin magnitude $<
0.989$ \cite{abbott16:_gw150}. The template grid used for the second
Advanced LIGO run has $\sim 400,000$ waveforms covering a parameter
space extending to larger masses, $M \lesssim 500 \Msun$, with a
denser sampling in the high-mass region with $q < 3$.

The template waveforms are obtained by solving for the coalescence
dynamics during the initial inspiral phase and the merger that follows
(for a review, see e.g., \cite{Hughes:2009iq}). Current targeted BBH
searches use template waveforms obtained from the quadrupolar $(\ell,
|m|)=(2,2)$ gravitational-wave modes emitted by binary of black holes
in quasi-circular orbits and with spins aligned to the orbital angular
momentum \cite{dal17:_desig_advan_ligos}.

This model is expected to be in good match with the gravitational wave
emission from binaries formed from isolated star progenitors in
galactic fields; but it ignores three effects that can be relevant in
other formation scenarios such as dynamical captures in dense stellar
environments. The effects so far poorly modelled and/or absent
in current searches are: higher-order than quadrupolar gravitational
wave modes, orbital precession due to non-aligned spins 
and orbital eccentricity.

The inclusion of those effects is not straightforward for several
reasons. First, gravitational wave modeling from binaries is still a
topic of active research and accurate waveform models are not always
available; for instance, this is true for binaries in eccentric
orbits \cite{Damour:2004bz,Tanay:2016zog,Moore:2016qxz}. Second, a
larger search space implies a larger template grid, thus larger
computing needs. For instance, the search for arbitrarily spinning (or
``precessing'') binary requires 10 times more template waveforms
(order of millions, using relaxed sampling density requirements). The
search algorithm has also to be adapted to account for the sky-dependency of the
signal received by the detector \cite{harry16:_searc_gravit_compac}.

Current searches based on the quadrupolar aligned-spin quasi-circular
template waveforms have partial sensitivity to signals that departs
from the nominal model. A fraction of the signals can be missed
because the signal \textit{vs} template phase agreement is not good
enough to obtain sufficient signal-to-noise ratio, or because the
signal is discarded by the ``chi-square'' consistency
test\footnote{This test measures how the amplitude profile of observed
signal in the frequency domain differs from that of the closest
phase-matching template.} \cite{Allen:2004gu,abbott16:_gw150} applied
to reject transient non-Gaussian instrumental noise. A number of
studies evaluate the ability of the current template grids to detect
signals departing from the quadrupolar aligned-spin quasi-circular
model.  Depending on the alternative astrophysical model (precession,
non-quadrupolar modes, eccentricity) being considered, $\sim$ 10 \% to
more than 50 \% of the sources can be missed
\cite{harry16:_searc_gravit_compac, calderon16:_impac,varma14:_gravit,
capano14:_impac,huerta17:_compl}.

%
%
%

Searching specifically for the aligned-spin quasi-circular quadrupolar
signals \cite{abbott16:_binar_black_hole_merger_first} can introduce
an observational bias as one more likely detect what one has searched
for \cite{harry16:_searc_gravit_compac}. The sources at or beyond the
boundaries of the currently searched parameter space may be of larger
interest as they are associated to somewhat unexpected astrophysical
scenarios.

Current matched filtering searches will extend in scope, e.g.,
by including more templates to cover a larger parameter space (see e.g.,
\cite{Indik:2016qky} for a proposal to cover precessing neutron-star
black hole binaries).

In the meanwhile, the unmodelled transient searches (see
e.g., \cite{abbott16:_obser_gw150}) provide an alternative approach
capable of identifying sources beyond the ones currently addressed by
matched filtering. Unmodelled transient searches rely on general
assumptions on the target signals instead of a precise model, and
detect the signal by identifying power excesses in a time-frequency
representation of the data.

The way the power excesses are arranged in the time-frequency plane is
directly related to the signal. In this paper, we propose to revisit
the pattern matching idea underlying in matched filtering searches,
and apply this idea to the time-frequency pattern rather than the
time- or frequency-domain waveform.  The goal is to improve the
sensitivity of existing burst searches by targeting given arrangements
or shapes motivated by astrophysical waveform models. This idea is
general but we apply it here specifically to one of the major search
algorithms referred to as coherent WaveBurst \cite{Klimenko:2015ypf},
used to analyze LIGO and Virgo data. Similar ideas have been explored
in \cite{Coughlin2015, Thrane:2014bma,coughlin14:_detec} in the context of other
data analysis pipelines, and in \cite{Suvorova+2016} in the context of
other searches.

The paper is organized as follows. In Sec.~\ref{SecII}, we briefly
introduce the coherent WaveBurst algorithm. Sec.~\ref{SecIII} shows
how astrophysical models can be encapsulated into a time-frequency
graph. This graph is used in a modification of coherent WaveBurst in
order to specialize the search to these particular models. In
Sec.~\ref{SecIV}, we apply this idea to the case of BBH signals and
evaluate the sensitivity of the new algorithm using simulated LIGO and
Virgo noise. The results are discussed in Sec.~\ref{SecV}.

\section{Overview of the Coherent WaveBurst pipeline}
\label{SecII}
Coherent WaveBurst (cWB) \cite{Klimenko:2015ypf} is a data analysis
pipeline used to search for gravitational wave transients with limited
prior knowledge on the waveform. The pipeline has been used to analyze
multiple LIGO and Virgo runs. In this section, we review the main
steps of cWB.

The data is mapped to the time-frequency domain by using the so-called
Wilson-Daubechies-Meyer (WDM) transforms defined
in \cite{Necula:2012zz}. The WDM transform provides a representation
of the data similar to short-term Fourier or Gabor analysis, but it
relies on an orthonormal basis composed of regularly distributed
sinusoidal functions that we later refer to as ``wavelets''. The time
and frequency resolutions of this representation are $M/f_s$ and
$f_s/(2M)$ respectively, determined by the chosen number $M$ of
frequency sub-bands and the sampling frequency $f_s$.
cWB computes a collection of WDM transforms over a range of
time/frequency resolutions to obtain a complete representation of
signal features that have different timescales. Typically, $M=2^\ell$ for
scales $\ell=4, \ldots, 10$ that correspond to analysis timescales ranging
from 7.8 ms to 0.5 s when $f_s=2048$ Hz.

From these collections of WDM transform cWB only retains the
time-frequency pixels that exceed a baseline amplitude which
corresponds to the last centile (or permille) under Gaussian noise
assumption.

Neighboring pixels in time and frequency are grouped in a
cluster $C$. In principle, clusters can have any shape. cWB applies only limits
cluster shape by fixing the gap between the pixels in time,
frequency, and time-frequency. They
provide a multi-resolution representation of the signal recorded by
the considered network of detectors.

Clusters are then characterized using a maximum likelihood ratio
statistics obtained from Gaussian noise assumption, assuming
the source is at the sky location $(\theta, \phi)$:
\begin{equation}
	\label{maxLratio}
	L_{max}(\theta, \phi) =\sum_{p \in C} \bm{w}_p^T \bm{P}_p  \bm{w}_p
\end{equation}

In the above equation, the sum runs over all pixels $p$ in cluster
$C$. The amplitudes (after whitening) of the time-frequency pixels are
collected into a vector $\bm{w}_p = \left\{
w_k(t_p-\tau_k(\theta, \phi), f_p,
M_p)/S^{1/2}_k(f_p)\right\}_{k=1,\ldots, K}$ where $\tau_k$ is the
time delay in the arrival time between the $k$th detector and a
fiducial reference point. The time $t_p$, frequency $f_p$ and scale
$M_p$ unequivocally characterize the time-frequency pixel $p$.

The operator $\bm{P}_p$ projects the data into the gravitational wave
subspace spanned by the noise-weighted antenna patterns $\bm{F}_+$ and
$\bm{F}_\times$, where $\bm{F}_+
= \left\{F_{k+}(\theta, \phi)/S^{1/2}_{k}(f_p)\right\}_{k=1,\ldots,
K}$ (and similarly for $\bm{F}_\times$) \cite{Klimenko:2015ypf}.  $K$
defines the number of detectors in the network and $S_k(f)$ is the
noise spectral power density for the $k$th detector.

The sum of the diagonal components of the quadratic form in the
summand of Eq.~(\ref{maxLratio}) defines the \textit{incoherent
energy} $E_{in}$, while the non-diagonal terms defines
the \textit{coherent energy}\footnote{The off-diagonal terms in
Eq.~(\ref{maxLratio}) do necessarily sum to a positive value for all
$\theta$ and $\phi$. The term ``coherent energy'' is not always proper.}
$E_{coh}$, i.e., from signal that are phase-coherent in all detector
observations \cite{Klimenko:2015ypf}.

The energy of the component in the data that does not lie in the
gravitational wave subspace is characterized by the \textit{null energy}
\cite{Klimenko:2015ypf,sutton10:_x_pipel}, viz.
\begin{equation}
	\label{nullenergy}
	E_{null}(\theta, \phi) =\sum_{p \in C} \bm{w}_p^T \bm{P}^{null}_p  \bm{w}_p
\end{equation}
where $\bm{P}^{null}_p = \bm{I}-\bm{P}_p$ is the projection operator orthogonal
to $\bm{P}_p$.

Those energies inferred from the data are combined into two statistics that
characterizes the amplitude and consistency of the signal associated
with the cluster. The network correlation coefficient $c_c =
E_{coh}/(|E_{coh}| + E_{null})$ allows to distinguish
gravitational-wave signals ($c_c \approx 1$) from spurious noise
events ($c_c \ll 1$)
\cite{Klimenko:2008fu,Klimenko:2015ypf}.
The statistic $\eta_c = (c_c E_{coh} K/(K-1))^{1/2}$ provides an estimate of the
network coherent signal-to-noise ratio
\cite{Klimenko:2015ypf,PhysRevLett.116.131103}.

Depending on the search being performed, other figure-of-merits are
 used together with the above two statistics. For instance, to improve
 background rejection in searches for compact binary coalescences, a
 selection cut on a crude chirp mass estimate has been
 introduced (see \cite{Tiwari:2015bda} for details).

\section{Basic principles of the proposed method}
\label{SecIII}
We now present a new clustering method called Wavegraph for the cWB
pipeline. In this section, we explain the general principles and
describe the major steps of the algorithm.

\subsection{General idea}

The proposed method combines three main ingredients. It is based on a
formulation of signal detection as pattern matching in the
time-frequency plane. An expected pattern is computed from the salient
time-frequency pixels i.e., pixels that stand above the noise level
when the signal is detectable.

As detailed in Sec~\ref{sec:sparse}, to compute this characteristic
set of pixels we employ an algorithm used for sparse signal
approximation that ensures this pixel set contains the complete
description of the waveform model.

Astrophysical scenarios generally provide a range of waveforms
parametrized by several physical source parameters. For instance,
compact binaries are characterized by the binary component masses and
spins. This leads to some variability in the expected time-frequency
pattern. We encapsulate the waveform model variability into a
time-frequency graph, see Sec~\ref{sec:graph}.

The graphical representation allows to formulate the detection problem
as a combinatorial optimization problem where efficient algorithms can
be used, as explained in Sec~\ref{sec:graph_search}.
In a nutshell, the algorithm seeks the cluster of time-frequency
pixels with the largest overall power (more precisely, the incoherent
energy carried by the cluster). While matched filtering techniques
perform a \textit{global} phase matching over the entire signal, thus
requiring a maximum accumulated phase difference between signal and
template lower than a fraction of a cycle, this type of search instead
performs a \textit{local} phase matching of the signal for each wavelet
associated to the pixels in the cluster. This can thus accomodate a
phase difference, resulting in more robustness at the expense of some
loss of efficiency.

In the following sections we detail all the steps mentioned above.
We start from set of characteristic gravitational signals for the
target astrophysical source. In the case of compact binary mergers, we
use a template bank employed in matched filtering based searches.

\subsection{Sparse signal representation}
\label{sec:sparse}

For each characteristic signal, the first step is to individuate a
representative set of salient time-frequency pixels. As mentioned
above, the cWB pipeline maps the data to the time-frequency plane
using a collection of WDM transforms, that is by projecting the data
onto a union of Wilson bases, thus resulting into an overcomplete
representation based on a redundant time-frequency dictionary.

Signal expansions in redundant dictionaries are not unique. Sparse
linear decompositions provide a complete signal representation where
the power is concentrated in a small set of dominating pixels, that
are thus more likely to stand above the noise level. Because of the
small number of pixels, there is a reduced chance that one or several
pixels in the decomposition match noisy transients that are often
observed in gravitational-wave data.

In an earlier version of the algorithm \cite{Chassande-Mottin:2017qim}
we proposed to compute the signal expansion from local maxima of the
WDM transforms. However, the resulting set of pixels generally failed
ensure full signal energy recovery, thus leading to an overall SNR
loss of 40 $\%$ on average.

The problem of sparse signal expansion in redundant dictionary has
received a good amount of attention over the last twenty years
\cite{mallat2009a}. Although this problem is NP-hard, efficient
algorithms are available.  The {\it matching pursuit} algorithm
\cite{Mallat:1993:MPT:2198030.2203996} is one of these and goes with
the following steps:
\begin{enumerate}
\item Decompose the signal onto the time-frequency dictionary,
\item Identify the dictionary wavelet with the largest dot-product,
\item Compute the residual signal by subtracting the contribution of the selected wavelet
  times the associated expansion coefficient.
\item Go to 1. until the residual energy is below a user-defined
  fraction of the original signal.
\end{enumerate}

\begin{figure}
	\hspace{-0.9cm}
	\includegraphics[width=1.1\linewidth]{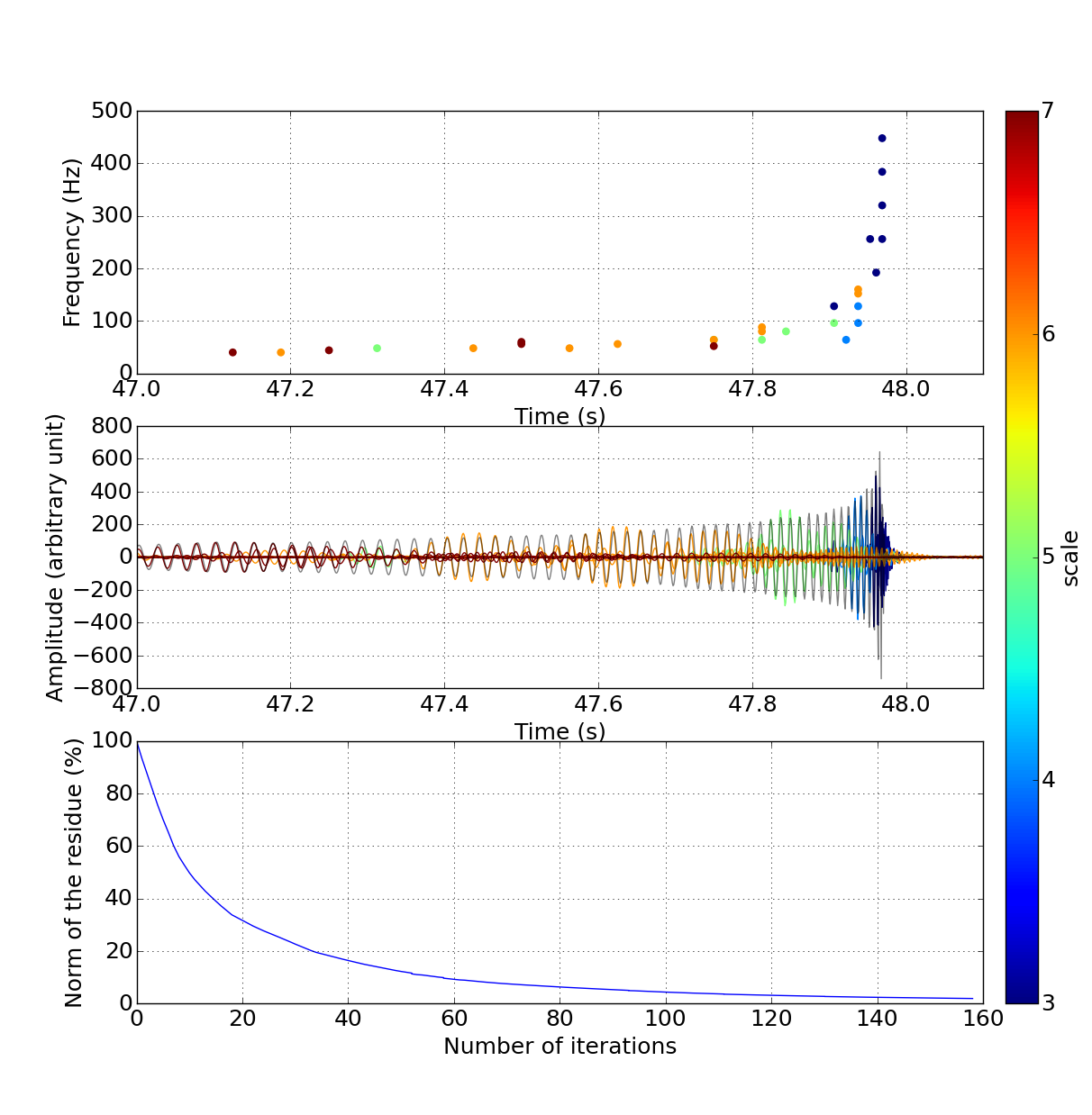}
	\caption{Convergence of the matching-pursuit algorithm in the case of 
	an equal-mass non-spinning BBH waveform with 20 $M_{\odot}$ total mass
	after whitening by advanced LIGO design sensitivity PSD. Fixing the
	termination condition at 20 \%, collects 33 pixels  and ensures
	sparsity (\textit{bottom}). Time-frequency pixels selected by 
	the matching pursuit algorithm (\textit{top}) and 
	corresponding wavelets (\textit{middle}) for the very same signal.
	The color indicates the scale $\ell$ associated to each pixel that
	corresponds to a wavelet timescales $2^\ell/f_s$. Strain amplitude of 
	the original waveform (\textit{black}) is shown in the background.
	Some early time pixels are not displayed that are selected by the
	matching pursuit.}
	\label{selectedpixelsWG}
\end{figure}

In this work, we set the fractional approximation error in the
termination condition to $\sim 20 \%$. After the first few tenths of
iterations (see the bottom panel of Fig.~\ref{selectedpixelsWG}), the
fractional error decreases slowly with the number of pixels selected
for the approximation\footnote{This is a well-known behaviour of the
  matching pursuit algorithm, see \cite{mallat2009a}.}. For instance,
choosing 10 \% instead of 20 \% error, the number of pixels almost
doubles. The extra signal power captured in the graph is distributed
over many low-amplitude pixels that only carry a tiny SNR fraction. In
presence of noise, those low-amplitude pixels are largely dominated by
noise. In our simulations, the waveform reconstructed from the cluster
extracted with the 10 \% error graph (see Sec.~\ref{sec:graph_search})
does not improve much, about 2 \% increase in SNR. We concluded that a
fractional error of 20 \% is a good compromise.

An example of the sparse decomposition obtained with the matching
pursuit algorithm is shown in Fig.~\ref{selectedpixelsWG}. It uses an
equal-mass non-spinning BBH signal with 20 $M_{\odot}$ total mass
(\texttt{SEOBNRv$2$\_DoubleSpin} model \cite{Purrer:2016}) after 
whitening by advanced LIGO design sensitivity curve. The top panel 
represents the selected time-frequency pixels evidencing the typical 
raising frequency evolution of gravitational-wave chirp signals.  
Closer to the final merger, the higher frequencies are captured 
by the smaller scales (shorter duration wavelets).

The middle panel displays the same signal in the time domain and the
approximation obtained by the collection of selected wavelets. The
bottom panel shows the norm of approximation error at each
iteration. As the number of iterations increase, the error decay
slower, thus requiring increasingly more pixels for each percent
improvement.


\subsection{Graph computation}
\label{sec:graph}

The set of time-frequency pixels selected by the sparse signal
approximation provides a compact representation of the waveform
model. With the selected pixels, we form a cluster by 
connecting pixels to keep a record of their co-occurrence
in association to the same signal. To limit the number of
interconnections we only connect pixels that are adjacent in a particular
ordering defined by a rule, e.g., by ascending order in time first,
and ascending in frequency in second. This thus forms a chain where a pixel
is connected to one pixel downstream, and another upstream.

We repeat this operation for different alignments in time between the
waveform and the coarsest time-frequency lattice associated to the WDM
transform with the largest analysis timescale. These alignments are
obtained by shifting the waveform in time by multiples of the smallest
analysis timescale. For the scale range $\ell=3, \ldots, 7$ used here, there
is thus a maximum of $2^4=16$ possible alignments.

We also repeat this operation on every waveforms associated with the
considered astrophysical scenario, i.e., the template bank.  
We thus obtain as many pixel clusters as there are templates in the
bank times the number of alignments.

The pixels in the clusters are characterized by the central time, frequency
and duration of their corresponding wavelet. The connections
between pixels form a time-frequency graph. As two
clusters may have pixels in common, pixels in the intersection receive
the connections they have in the two clusters of origin. They thus end
up being connected to more than one pixel downstream or upstream.  The
graph is oriented and acyclic, for well chosen pixel ordering rule.

The resulting graph provides a compact and convenient representation of
the entire waveform manifold associated with an astrophysical
scenario. It allows to use efficient search algorithms borrowed from
graph-based combinatorial optimization as we will see in the next
section.

We provide examples of graphs produced for BBH searches in Sec.~\ref{SecIV}
(see Fig.~\ref{fig:graph_R1_withTS}).

\subsection{Search observational data using the graph}
\label{sec:graph_search}

Signals from the targeted family can be searched by finding the cluster $C^{\star}$ in
the graph that maximizes the likelihood ratio in
Eq.~(\ref{maxLratio}). However this would require the computation of
$L_{max}$ over the sky coordinates which is too computationally
expensive. Instead we identify the interconnected cluster in the
graph $G$ that satisfies
\begin{equation}
  \label{eq:rankingstat}
  C^{\star} = \argmax_{C \in G} \sum_{p \in C}  E_p - \lambda \bar{E}(f_p, M_p),
\end{equation}
where $E_p = \sum_k \hat{w}^2_{p,k}$ with
\begin{equation}
\hat{w}^2_{p,k} = \max_{\theta,
  \phi} w^2_k(t_p-\tau_k(\theta, \phi), f_p, M_p)
\end{equation}

The first part of the sum in (\ref{eq:rankingstat}) is a proxy for the
likelihood ratio in (\ref{maxLratio}). Though not identical, the term
$E_p$ is related to the incoherent energy $E_{in}$ introduced earlier
in Sec.~\ref{SecII}. $\bar{E}(f_p, M_p)$ is an estimate of the noise
level at a given $f_p$ and $M_p$ obtained by the median value of $E_p$
at this frequency and scale for all times.

The second term of the summation in (\ref{eq:rankingstat}) is a kind
of ``Occam's razor'' penalization term that favors small
clusters. Without this penalization, the maximization in
(\ref{eq:rankingstat}) would have the tendency to prefer large
clusters that accumulate more noise power than smaller signal-related
clusters.  The factor $\lambda$ allows to tune the strength of this
penalization. Typically, we set $\lambda=1$.

Maximizing Eq.~(\ref{eq:rankingstat}) can be directly related to the class of {\it "longest path problem"} and can be solved by the {\it dynamic
programming} algorithm. The computational cost scales linearly with the number of connections in the graph. This makes it
particularly efficient even for complex graphs. 

When applying the search to a segment of observational data, this
segment is divided into successive blocks shifted by strides
equivalent to the largest analysis timescale. The best-fit cluster
$C^{\star}$ is computed in all blocks.

The follow-up of the selected clusters is performed following the
standard cWB workflow, by computing their coherent statistics $c_c$
and $\eta_c$ given in Sec.~\ref{SecII}. We measured that the use of
this clustering algorithm in cWB results in an executation time about
20 \% larger compared to standard cWB on the same computer.

\section{Results}
\label{SecIV}
While this method is applicable in principle to a broad range of
astrophysical models, we will use the specific case of compact binary
mergers to demonstrate the idea.  In this section we show how
wavegraph can improve upon cWB in searching for BBH signals in
simulated Gaussian noise. We give a comparison with the recent version
of cWB that was used in \cite{abbot17:_2017PhRvL.118v1101A}.

\subsection{Time-frequency graphs}

We carry out simulations of a BBH search using cWB with and without
Wavegraph. In order to check the effect of using graphs with different
size and complexity we divided the mass range in two disjoint regions.

The region R1 (low-mass) corresponds to a total mass range of
$10$ - $25$ $\textsc{M}_{\odot}$, while R2 (higher-mass) covers $40$ -
$70$ $\textsc{M}_{\odot}$. The selected range of masses is similar to
that of the BBH events observed during the first two LIGO
observational runs. For both regions, we set the mass ratio
$q=m_2/m_1 \leq 2$ and cover the entire spin range available for the
waveform model
\texttt{SEOBNRv$2$\_ROM\_DoubleSpin}
\cite{Taracchini:2012ig,Taracchini:2013rva,Purrer:2015tud}
up to almost maximally spinning black holes. 

Using the algorithm described in Sec.~\ref{sec:graph} we compute the
time-frequency graphs using grids of template waveforms that covers
the mass range chosen for each regions.

We used template grids of 28\:201 (resp. 2\:950) waveforms for the R1
(resp. R2) region computed using a stochastic placement
algorithm \cite{Harry:2009ea}. To limit the overall computational cost
for the time-frequency graph, the set of template waveforms is
produced with a slightly coarser sampling of the parameter space in
the R1 case (minimal match of 95\% for the R1 graph, and 99\% for R2)
and used a limited number of waveform alignments; only 1 for R1, and
32 for R2.

This leads to a graph with 1643 pixel nodes for R1, and 941 nodes for
R2, a difference that can be explained by the relatively shorter signal
duration for higher-mass binaries in R2.

\begin{figure*}
	\vspace{-1cm}
	\begin{minipage}[b]{\linewidth}
		\hspace{-2cm}
		\includegraphics[scale=0.5]{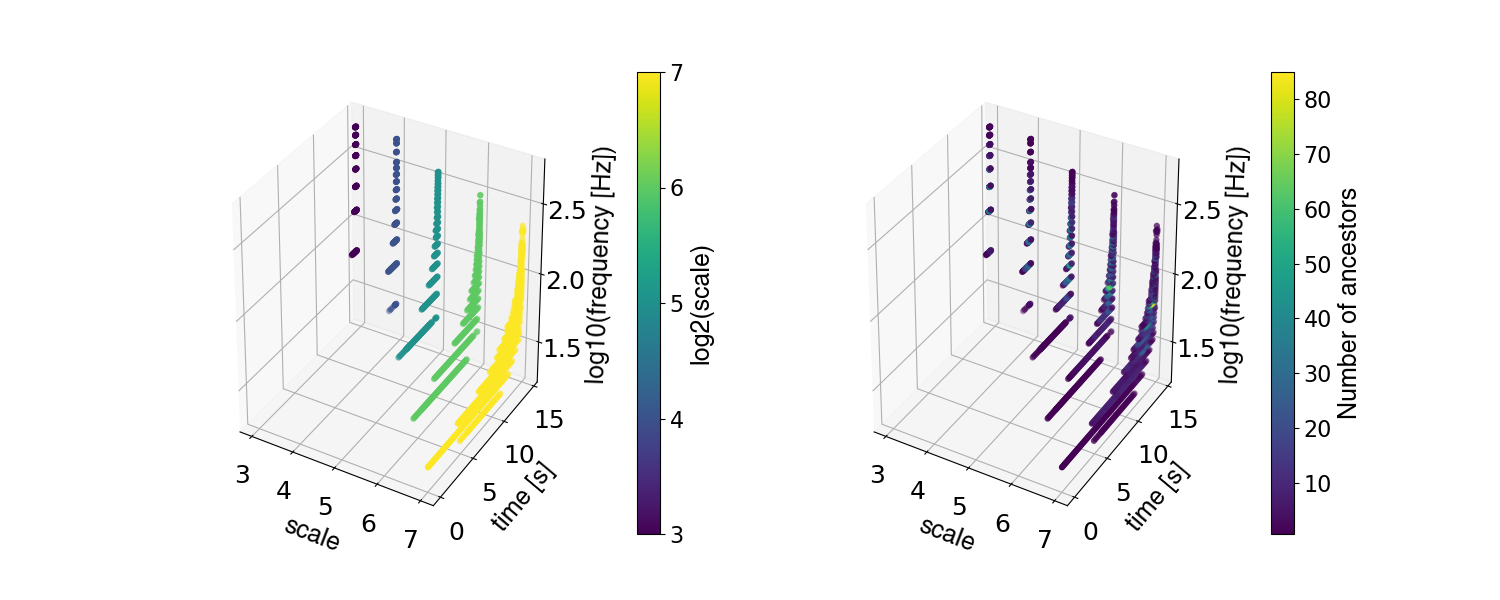}
	\end{minipage} 
	\vfill
	\vspace{-0.3cm}
	\begin{minipage}[b]{\linewidth}
		\hspace{-2cm}
		\includegraphics[scale=0.5]{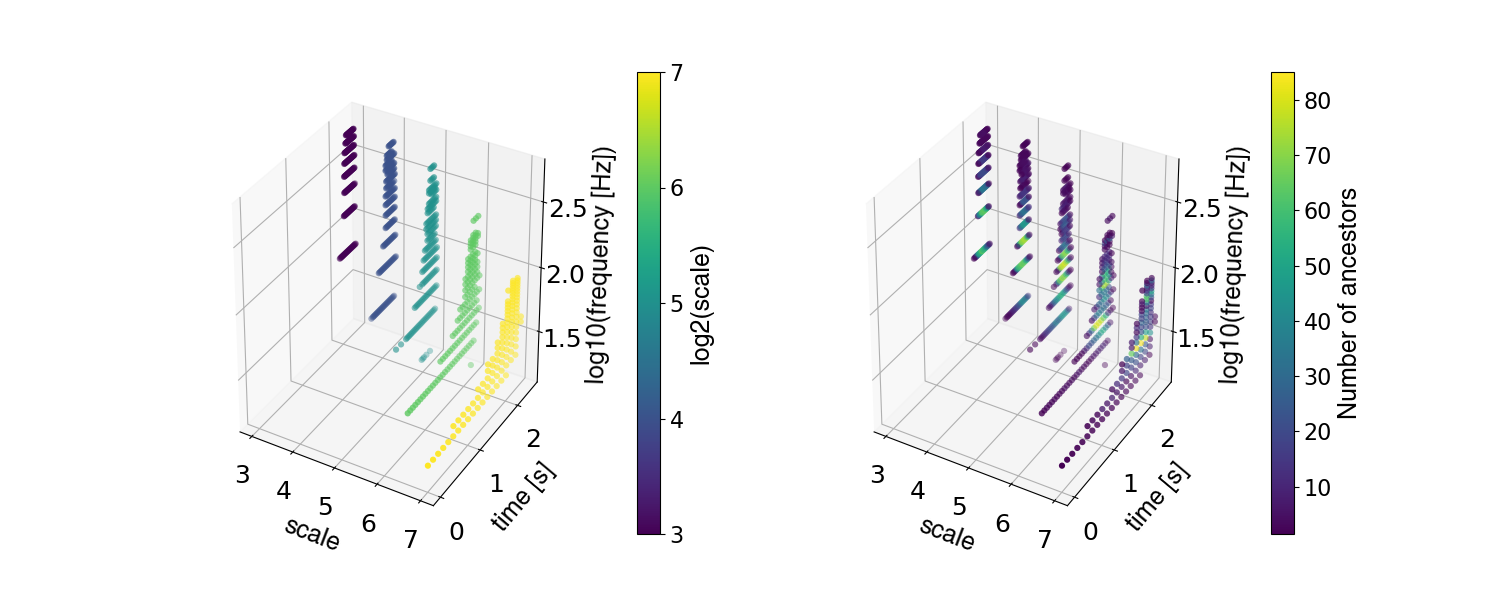}
	\end{minipage}
	\caption{Time-frequency graphs used to search for BBH signals with the wavegraph algorithm.
	R1 region (\textit{top}) and R2 (\textit{bottom}). The number of ancestors refers
	to the number of edges linking a given pixel to its neighbours that precede that pixel in the selected pixel ordering. It is an indicator of 
	the graph complexity.}
	\label{fig:graph_R1_withTS}
\end{figure*}

Top left panel of Figure \ref{fig:graph_R1_withTS} displays the pixels
in the graph computed for R1 and R2 with time on $x$-axis, frequency
on the $y$-axis and scale encoded in colors. The panel on the right
shows the graph connectivity \textit{i.e.}, the number of connections
of a given time-frequency pixel with its neighbors.

As expected, the overall graph time support and complexity is larger
for R1 as the BBH signals have a longer duration for this mass range
and there is a higher density of template waveforms.

We thus perform two searches using cWB with Wavegraph and the R1 and
R2 graphs separately. We do not intent to combine the two searches a
posteriori (which would then require to apply a trial factor to
account for the multiple search attempts). The idea is rather to show
how the search background changes with the different graphs.

\subsection{Simulated data set}

We present a Monte-Carlo simulation of the three detector network
composed of the two LIGO and the Virgo detectors. We used simulated
Gaussian noise colored according to the advanced LIGO and Virgo design
sensitivities given in \cite{aasi15:_advan_ligo, acernese15:_advan_virgo}.

We simulate and add about $10^6$ signals from binaries that are
arbitrarily oriented, isotropically and uniformly distributed in
volume up to maximum (luminosity) distances of $1.4$ Gpc and $3$ Gpc
for the R1 and R2 analysis respectively. These limits are much larger
than the search distance reach. The source sample thus includes many
low SNR binaries, a majority of which are below the detection
threshold.  This allows to correctly estimate the sensitive distance.

Cosmological effects are ignored in this simulation. The binary
population is thus not exactly distributed uniformly in comoving
volume. We expect that this to have little impact on the final result
as we are primarily interested in the \textit{relative} sensitivity
improvement between pipelines and the bulk of the detected
sources is located at low-$z < 0.1$.

The mass ranges of the simulated signals are consistent 
though slightly larger than the span used to compute the time-frequency 
graph.  A small fraction of the injected signals may then fall outside 
of the time-frequency graph mass coverage.





\subsection{Results}

In this section, we summarize the main results obtained running the Monte-Carlo simulation presented in the previous section.


\subsubsection{Background estimation}

We compare the search algorithm sensitivities at a fixed false alarm
rate (FAR). We evaluate the false-alarm rate by generating 129 days of
simulated Gaussian noise for the three detectors. We then follow the
background estimation procedure classically used for actual searches. It is based on
surrogate data produced by applying nonphysical time-shifts (larger
than time-of-flight between detectors) \cite{PhysRevLett.116.131103}.
By applying this procedure to the original data set with $\sim 600$
time lags, we generate the equivalent of 212 years of time-shifted
surrogate data. The analysis of this noise-only data results in the
search background, \textit{i.e.} a set of noise-related events each
characterized by their statistics $\eta_c$ and $c_c$ defined in
Sec.~\ref{SecII}. The search background is further used to associate a
FAR estimate to any detected signals by counting the number of
background events passing the same selection cuts. The FAR estimate is
a measure of the event significance, \textit{i.e.}, the probability
for an event to come from the noise.

We apply selection cuts similar to that used in actual searches.
These include the ones introduced for the rejection of transient
noise (or glitches) present in the LIGO and Virgo data, although this
type of noise is not present in the simulated data used for this
study. Signals with a low network correlation coefficient $c_c$
are discarded (poor phase coherence between pairs of detectors). We have used $c_c=0.7$ for both analysis.

Clusters of pixels associated with only few
wavelet durations (or time-frequency resolutions) are also discarded
as this type of cluster is not representative of what is expected for
BBH signals (see Sec.~\ref{sec:sparse} and
Fig.~\ref{selectedpixelsWG}). The cWB analysis includes a third
selection cut based on a chirp mass estimate obtained from the
reconstructed time-frequency pattern (see \cite{abbott16:_obser_gw150}
for details). This selection retains clusters that possess a
``chirping-up'' time-frequency signature compatible with BBH signals.


\begin{figure}[h!]
\centering
         \includegraphics[scale=0.3]{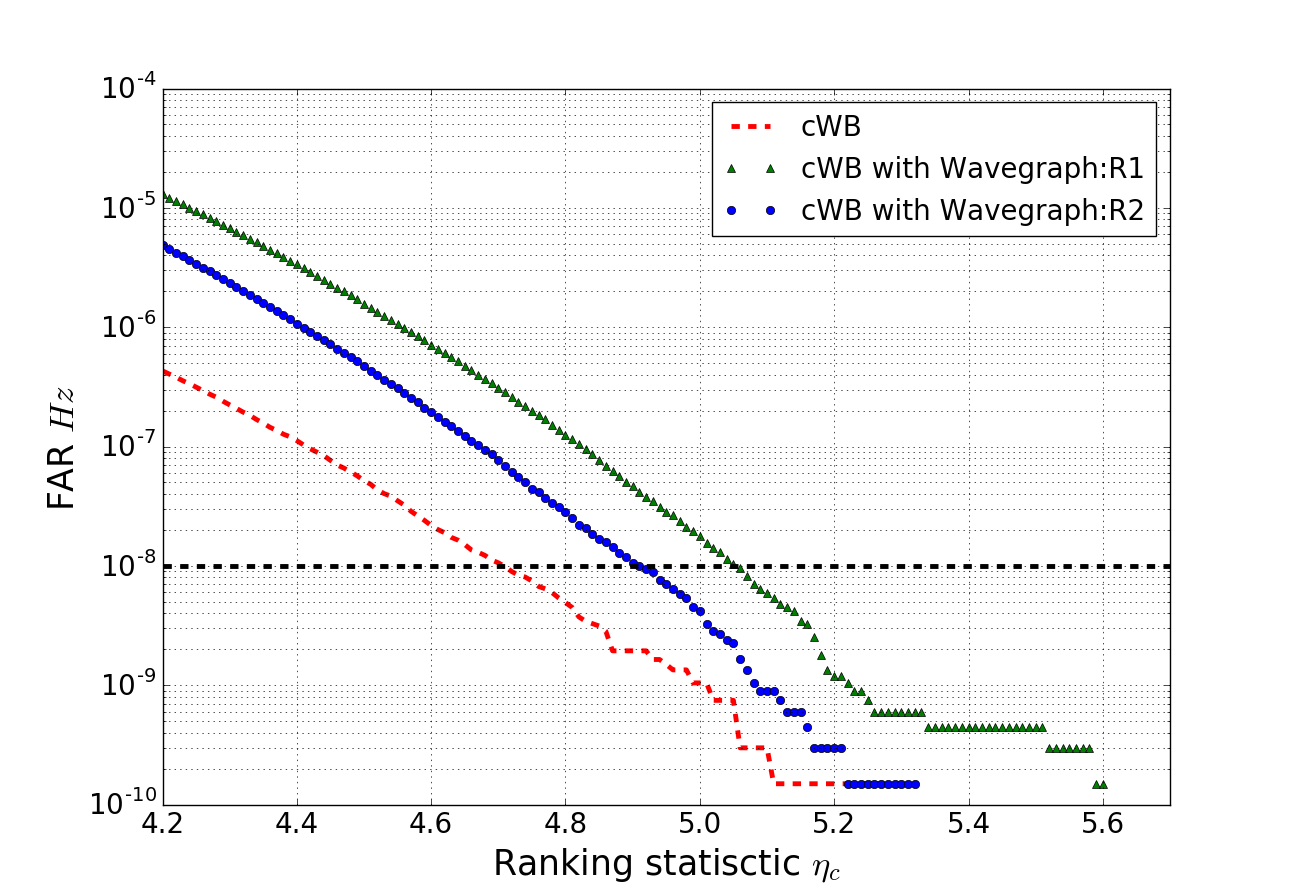}
         \caption{
         Cumulative distribution of background events \textit{vs} the
         statistic $\eta_c$ for cWB red/dashed line, with cWB with wavegraph using R1
         graph green/triangle line and using the R2 graph in
         blue/square line. Dotted line shows the reference FAR of
         $10^{-8}$ Hz = $0.3$ event/yr, chosen to compare the different
         pipelines.}  \label{bkgcurves}
\end{figure}

Figure \ref{bkgcurves} represents the cumulative distribution of the
background as a function of the ranking statistic $\eta_c$.  Both
analyses using wavegraph have higher background rate than cWB alone.
Our explanation is that the clusters extracted using wavegraph are, by
design, larger on average and with a wider spread in time. Also, while
cWB gathers sets of contiguous pixels with large amplitudes, wavegraph
allows for interruption, since the cluster is selected whenever the
overall energy evaluated globally is large whether the cluster is a
connected set or not. These two effects have the consequence to
slightly expand the signal space accessible to wavegraph.

The analysis using the R1 graph has a higher background compared to
R2. We relate this to the increase in size and complexity of the graph,
that has about twice more nodes in the R1 case, which thus increases the
probability of picking up noise outlier.

At the reference FAR=$10^{-8}$ Hz ($0.3$ event/yr) adopted here, we
obtain the $\eta_c$ selection threshold of 4.7, 4.9 (+4 \%) and 5.05
(+5 \%) for cWB, cWB with wavegraph (R1) and cWB with wavegraph (R2) respectively.  We
estimate the accuracy on the determination of those thresholds to be
$\lesssim 0.6 \%$. In the following, we apply this selection
threshold on $\eta_c$ to declare a signal detected.

Since this ranking statistic is homogeneous to the signal-to-noise
ratio and is thus inversely proportional to the distance, we may
conclude that one loses few percents in distance reach by using
wavegraph. This conclusion is however not correct as it does not fold
in the amount of SNR recovered on average in presence of a real signal
as we will see in the next section.

\subsubsection{Signal recovery}

In this section we examine the average properties of the BBH signals
from the injection population that are detected by cWB and cWB with
wavegraph.

\begin{table}
\begin{tabular}{|>{\raggedright}p{1.6cm}|>{\raggedleft}p{1.4cm}|>{\raggedleft}p{1.4cm}|>{\raggedleft}p{1.4cm}|>{\raggedleft\arraybackslash}p{1.4cm}|}
\hline
\multicolumn{1}{|c|}{Algorithm} &
\multicolumn{2}{|c|}{R1 region} &
\multicolumn{2}{|c|}{R2 region} \\
& \centering Injected & \centering Recovered  & \centering Injected & Recovered \\
\hline
cWB & \raggedleft 930\,744 & 28\,900 (3.1 \%)& 930\,870 & 26\,927 (2.9 \%)\\
\hline
cWB with wavegraph & 930\,744 & 35\,340 (3.8 \%) & 930\,870 & 33\,699 (3.6 \%)\\
\hline
\end{tabular}
	\caption{Number of injected/recovered signals by the simulation in the R1 and
	R2 regions for the two compared pipelines. Note the recovered signals indicated in the table are not exclusive to a single pipeline.}
	\label{table1}
\end{table}	

Table \ref{table1} tabulates the summary of injected and recovered BBH
signals in both R1 and R2 regions by cWB with and without wavegraph.
The numbers show the percentage improvement in recovery
with and without wavegraph. About $\sim 35-40 \%$ of the events
recovered with wavegraph are missed by cWB alone (exactly, $14\,627$
and $11\,402$ for R1 and R2 regions respectively). Conversely,
$15-25 \%$ of the events recovered by cWB alone are missed when using
wavegraph (exactly, $7\,941$ and $4\,492$ for R1 and R2 regions
respectively). There is thus a good complementarity.

\begin{figure}[h!]
        \centering
        R1 (low-mass) search\\
        \includegraphics[width=0.5\textwidth]{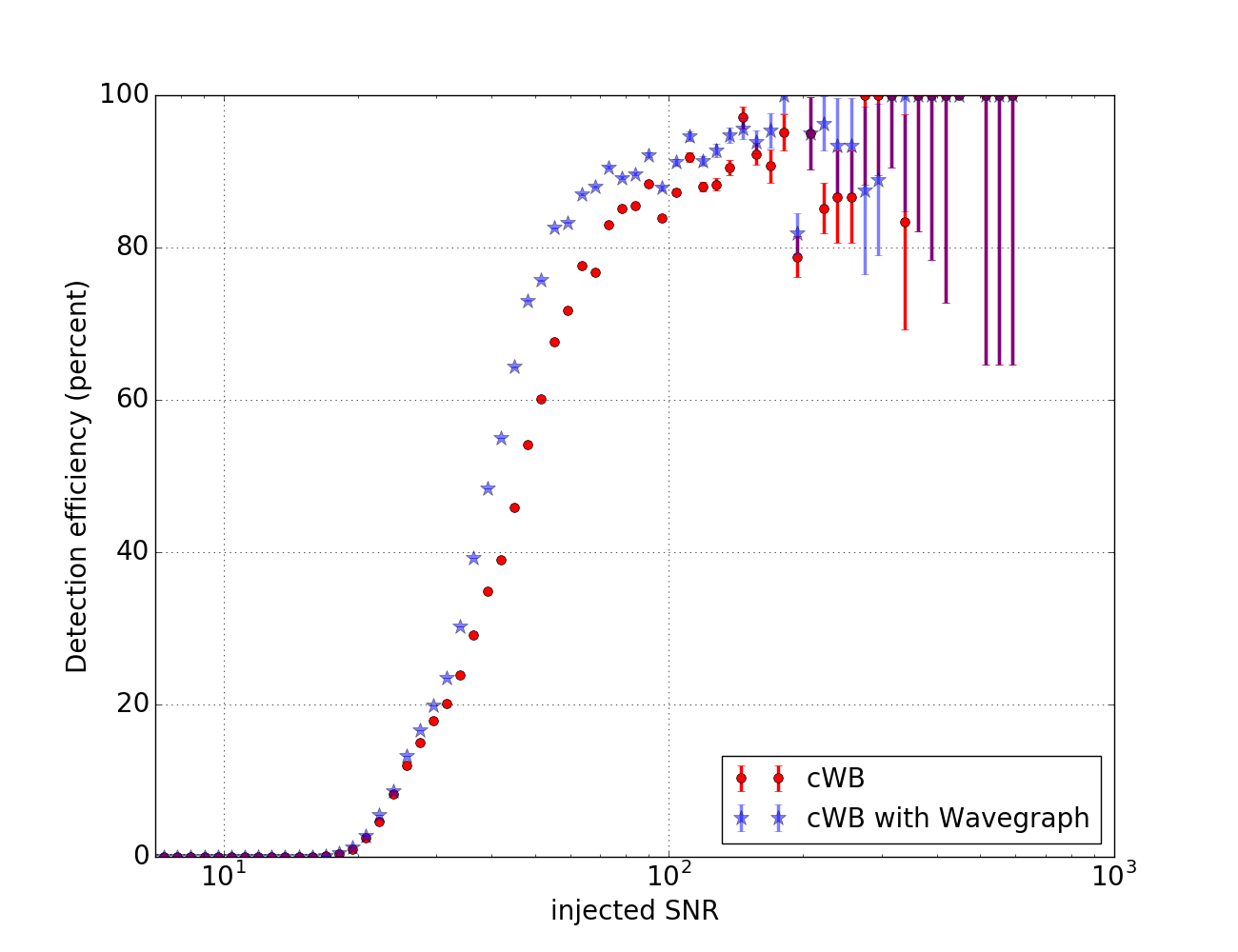}
        R2 (higher-mass) search\\
        \includegraphics[width=0.5\textwidth]{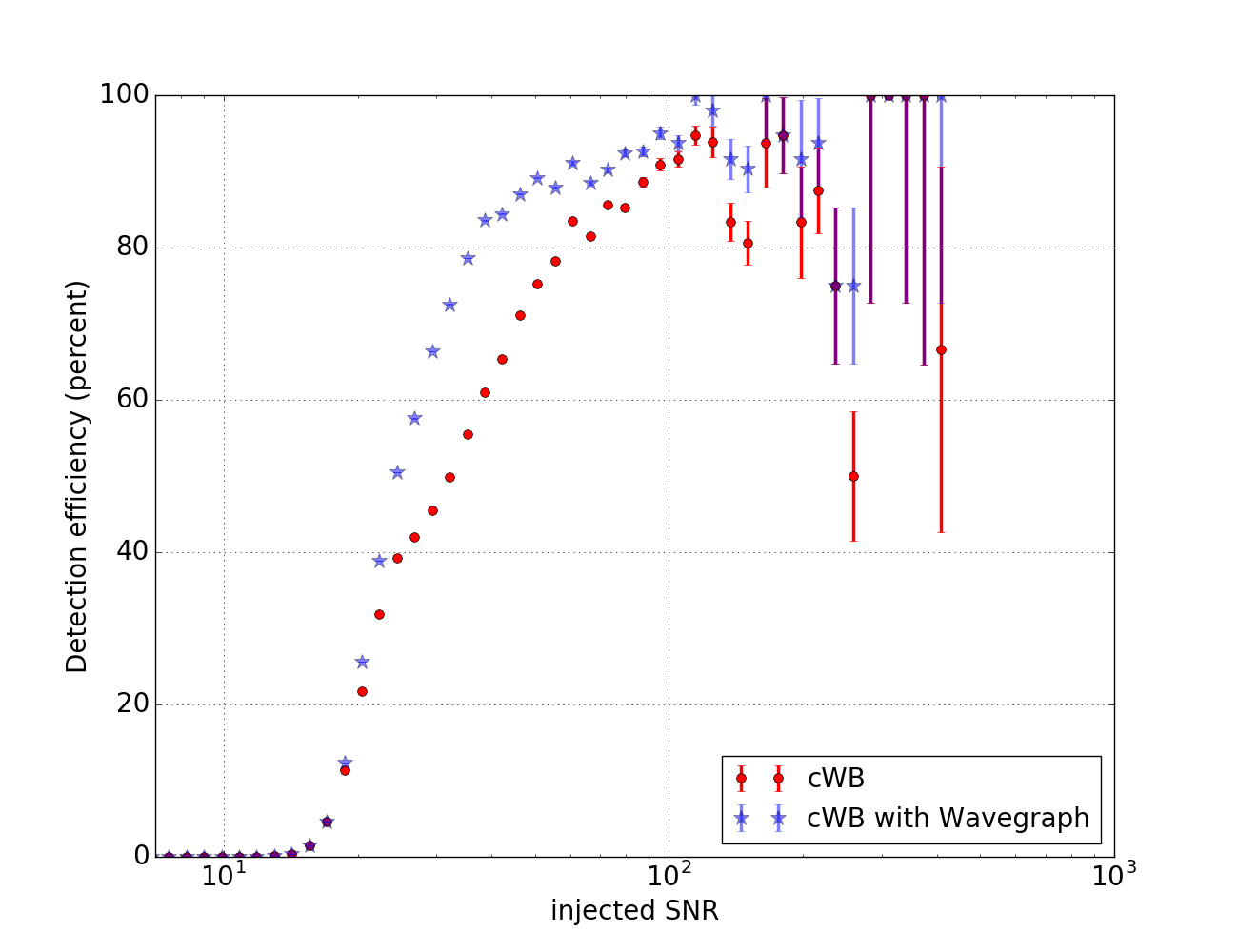}
\caption{Detection efficiency (percent) \textit{vs.} the injected network SNR 
        given for cWB in red circles and for cWB with wavegraph in
        blue stars with 1-sigma error bars. The top and bottom panels
        correspond to R1 and R2 simulations
        respectively.}  \label{recoveredevent}
\end{figure}

\begin{figure}[!ht]
        R1 (low-mass) search\\
   	\includegraphics[width=0.5\textwidth]{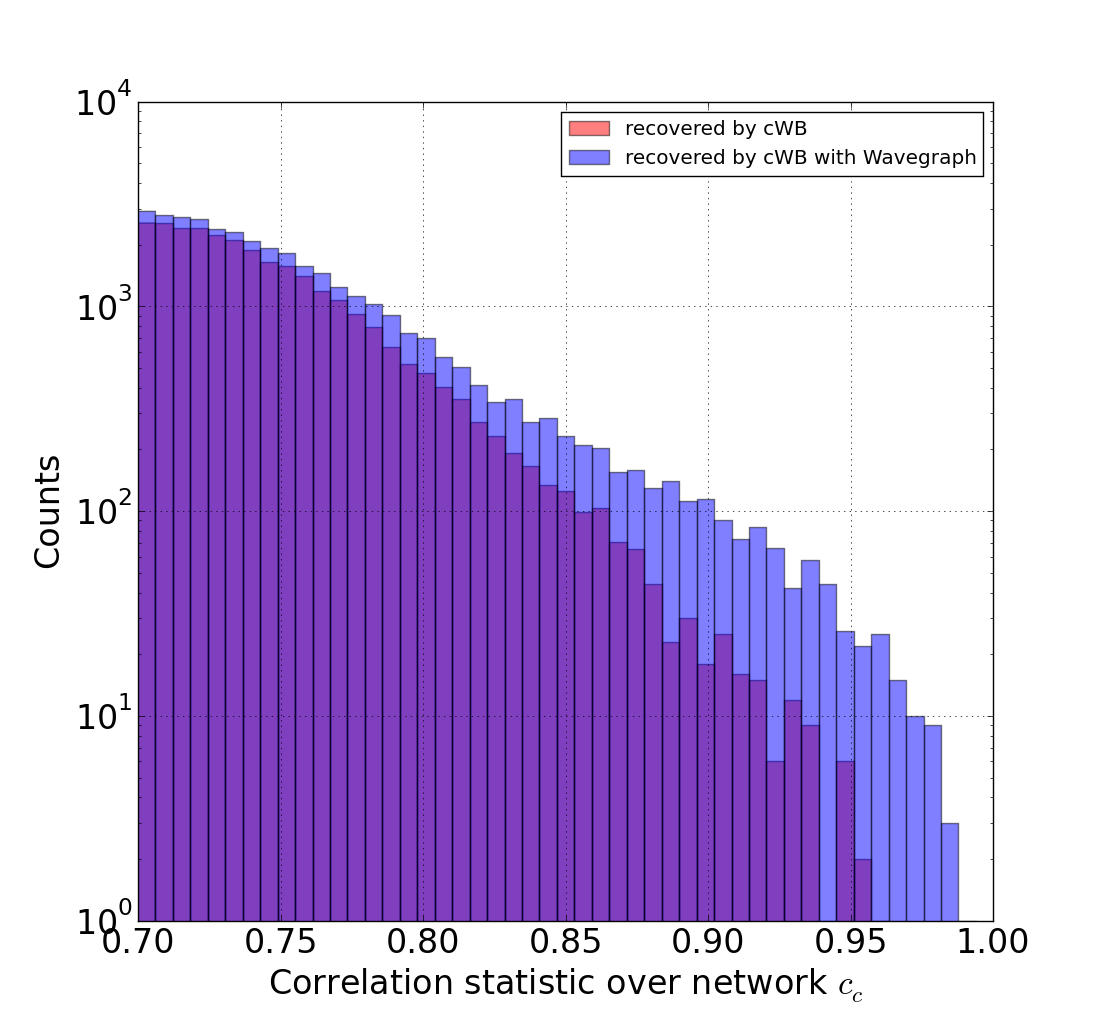}
        R2 (higher-mass) search\\
   	\includegraphics[width=0.5\textwidth]{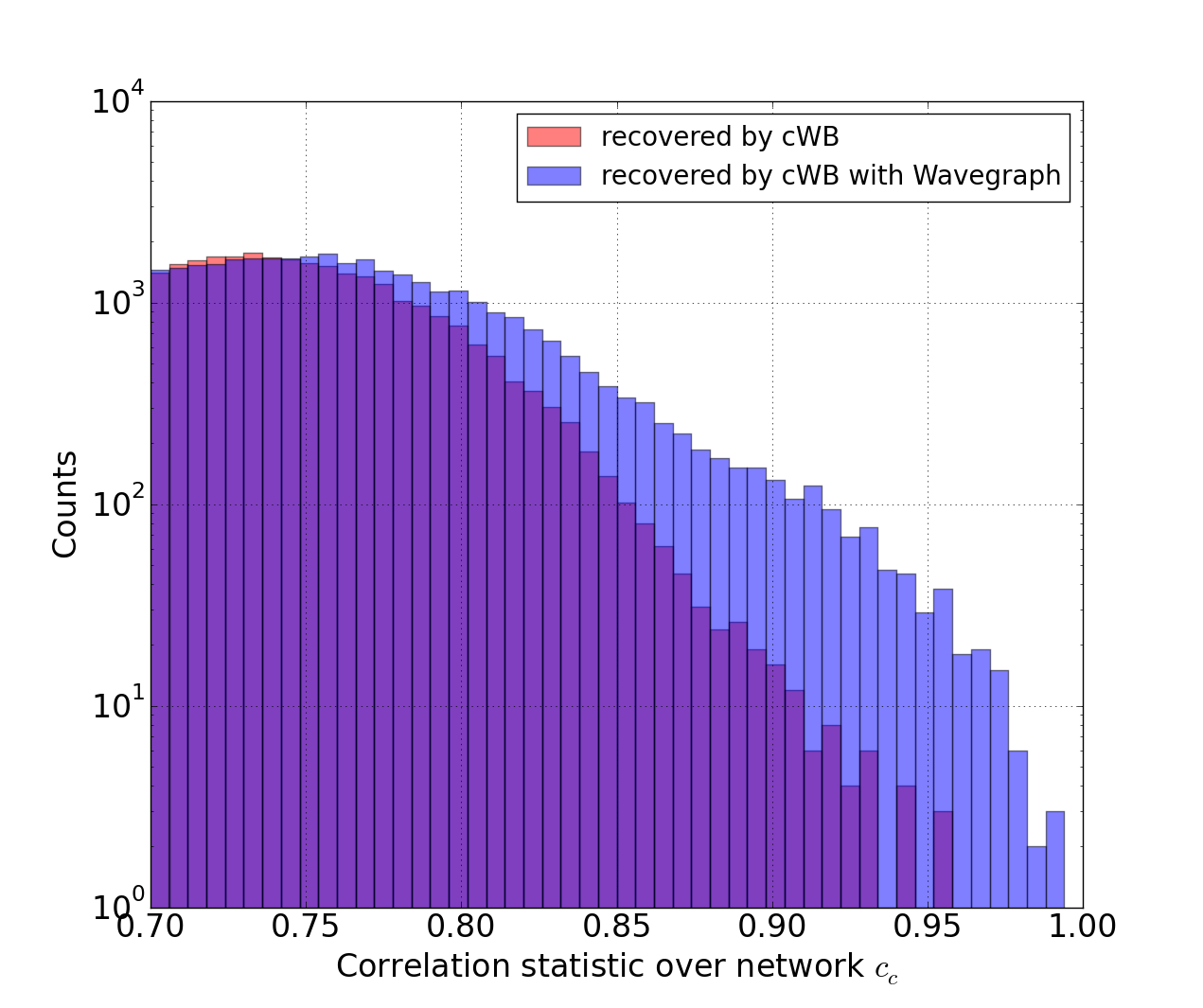}
   	\caption{Histograms of the network correlation coefficient of
   	recovered events by the R1 (\textit{top panel}) and R2
   	(\textit{bottom panel}) simulations. Both show the counts for cWB (\textit{red/darker}) and cWB with wavegraph (\textit{blue/lighter}).
        Wavegraph is on average reconstructing events
   	with a greater correlation over the network. The effect tends
   	to be even more pronounced with louder
   	injections.}  \label{histnetcc}
\end{figure}

In Fig.~\ref{recoveredevent} we display the detection efficiency
(percent) \textit{vs.} the injected network SNR for cWB and cWB with wavegraph.
For the two mass regions, the results shows that
the use of wavegraph improves the detection efficiency especially in
the low injected network SNR region, though we applied a more
selective threshold on $\eta_c$ to keep the FAR requirement equal for
all searches. 

In Figure~\ref{histnetcc} we show the distribution of the network
correlation $c_c$ statistic for the signals recovered by cWB
(\textit{red/darker}) and with wavegraph (\textit{blue/lighter}). It
appears that using wavegraph, the recovered signals have an higher
$c_c$ statistic for both R1 and R2 searches. Thank to the information
extracted from the waveform model stored in the time-frequency graph,
the wavegraph algorithm is more likely to pick pixels relevant to the
gravitational-wave signal, leading to an overall larger correlation.

We estimate the distance reach by computing the sensitivity distance
within which we would be able to detect a signal, averaged over the
observation time and over source sky location and
orientation \cite{2013PhRvD..87b2002A}. We checked that this distance
scales as expected with $\mathcal{M}^{5/6}$ where $\mathcal{M}=(m_1
m_2)^{3/5}/(m_1+m_2)^{1/5}$ is the chirp mass.

\begin{figure}[!h]
\includegraphics[scale=0.3]{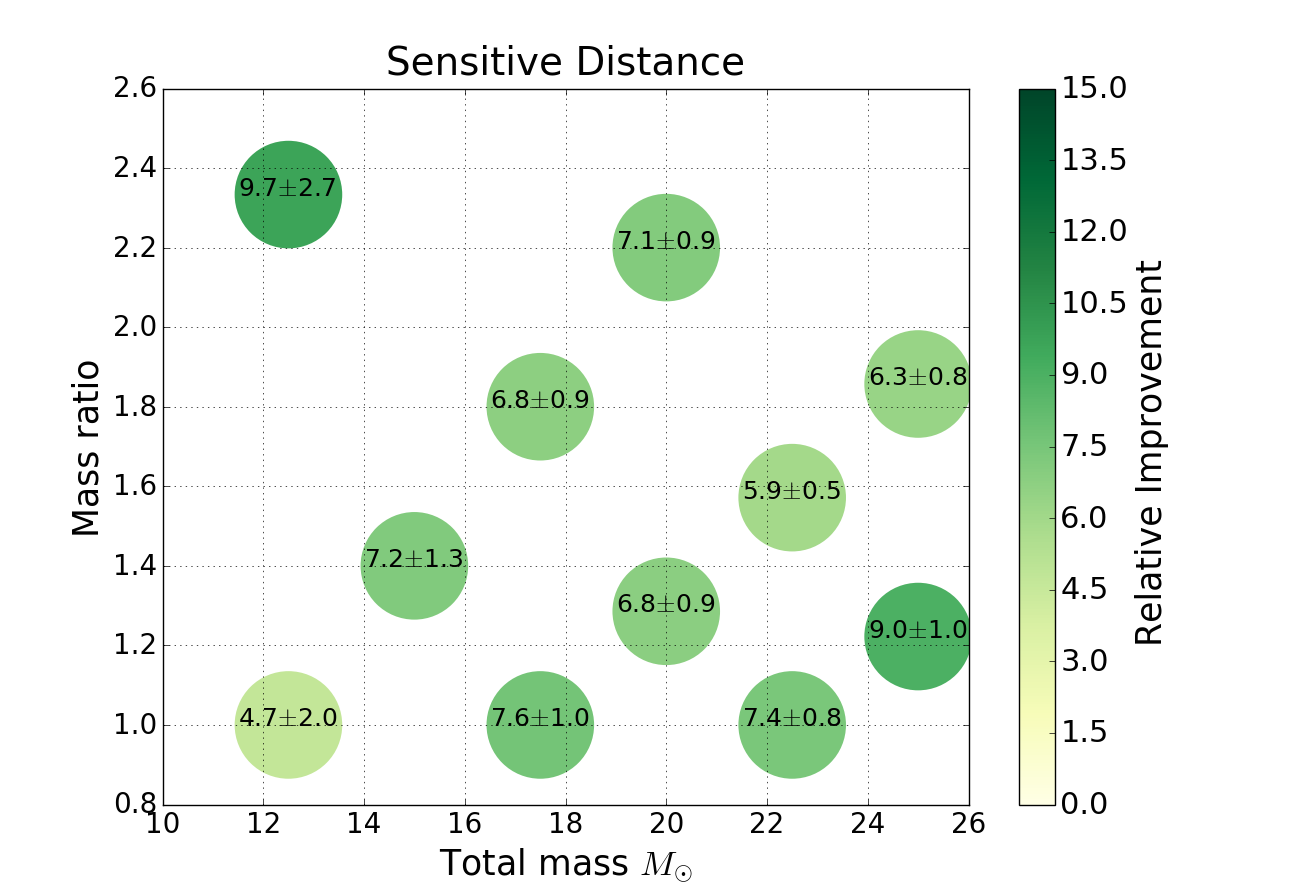}
	\includegraphics[scale=0.3]{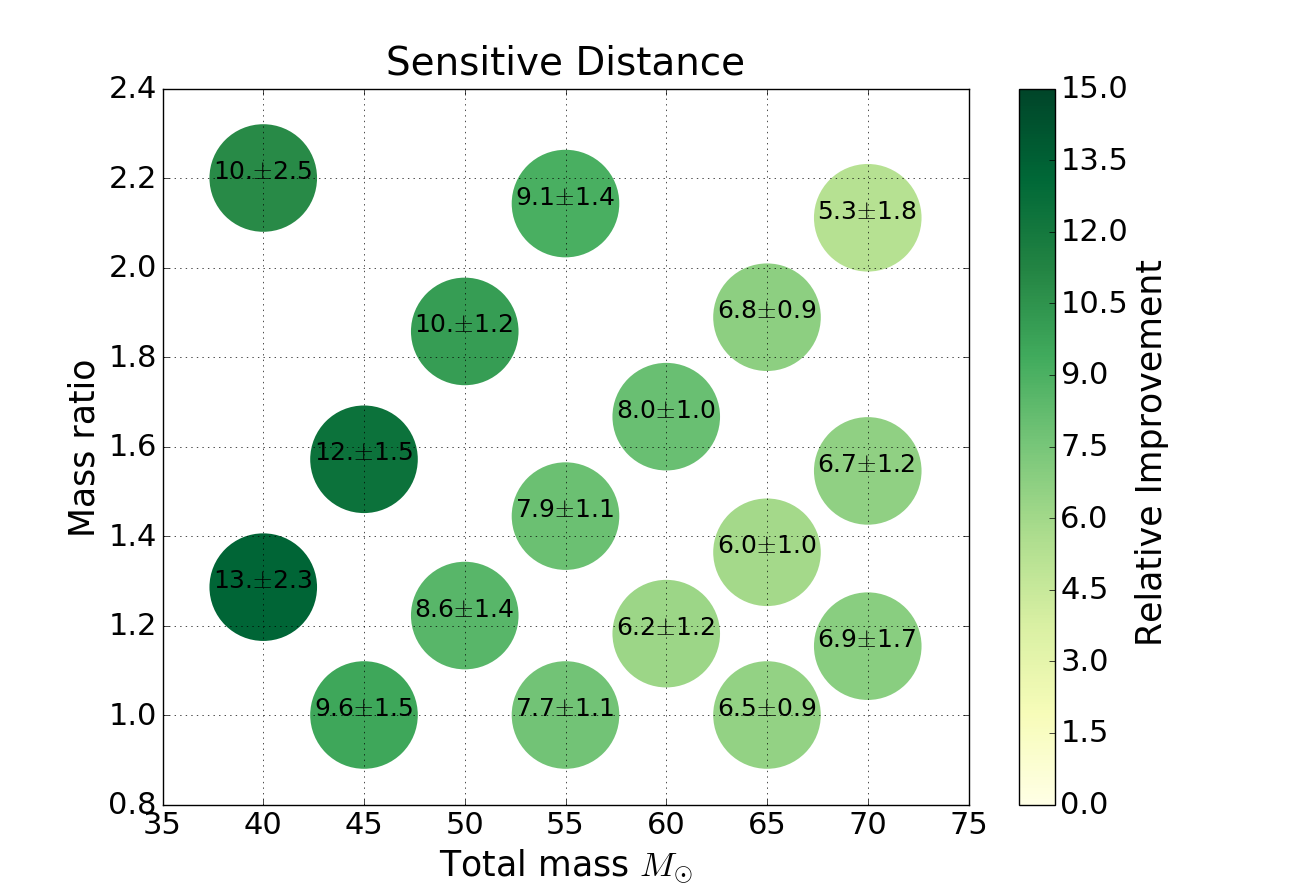}
\caption{Relative sensitive distance improvement (in percent)
when using cWB with wavegraph with respect to cWB alone for R1
(\textit{top panel}) and R2 (\textit{bottom panel}).
The color scale encodes the importance of the
improvement with respect to cWB alone}
\label{effectivedistancetiles}
\end{figure}

Figure \ref{effectivedistancetiles} displays the relative improvement
in the sensitive distance for various mass bins for R1 (top) and R2
(bottom) regions. In the R1 region, the average relative improvement
in sensitive distance when using cWB with wavegraph is $\sim 7$\% with a
maximum of $9.7 \%$ for asymmetric binaries ($q \sim 2$). In the R2
region, the average improvement is $\sim 8 \%$ with a maximum at
$13 \%$ in the lower part of this mass range. Overall, this translates
into an improvement in the event of about $20-25 \%$ at a FAR level of
$0.3$/yr. We observe that the improvement in the sensitive distance
decreases with the total mass for a fixed mass ratio. As the total
mass increases, the BBH chirp signal shortens and cWB is able to
collect all relevant pixels.  The dependencies of the relative
sensitive distance improvement with the mass ratio, for a fixed total
mass is rather weak.

\section{Conclusions}
\label{SecV}
In this work we show that it is possible to improve the sensitivity of
the time-frequency pattern search algorithms used in the context of
unmodelled gravitational-wave transient searches, such as cWB, by
restricting the exploration to patterns obtained from
astrophysical waveform models.

We propose a pattern matching algorithm called ``Wavegraph'' that we
implemented as a module in the cWB search pipeline. The algorithm has
no significant impact on the overall computational cost of the search.

We applied and tested this algorithm on the case of BBH
signals, and this leads to an averaged 7-8\% improvement in the
standard cWB distance reach over the considered mass range (total mass
of $10-70 M_{\odot}$). This translates into a 20-25 \% relative
improvement in the detection rate, assuming an isotropic source
distribution. We attribute this enhancement to the more efficient
collection of signal-related time-frequency pixels achieved by the
proposed algorithm. Our tests evidence the impact of the graph size
on the background level and search sensitivity.

The method is an intermediate approach between matched-filtering
based searches that rely on the precise knowledge of the signal phase
evolution and unmodelled searches that impose little priors.  We
consider the method particularly adapted to areas of the compact
binary space such as high-mass ratio binary and precessing or
eccentric orbital evolution, where it is difficult to build an
exhaustive template bank thus preventing standard modelled searches to
operate.

This method is general and applies to a broad range of situations,
including the cases where only numerical waveform models (e.g., from
numerical relativity simulations) are available.

Similarly to cWB, the Wavegraph algorithm is sensitive to instrumental
and/or environmental transient noise. In a follow-up article, we
assess the impact of real detector noise on the search and develop
additional glitch rejection scheme allowed by the time-frequency
graph to circumvent this issue \cite{secondWavegraphPaper}.

\section{Acknowledgements}
This research was supported by the CEFIPRA grant No IFC/5404, by the
European Union's Horizon 2020 research and innovation programme under
grant agreement No 653477, by the French Agence Nationale de la
Recherche (ANR) under reference ANR-15-CE23-0016 (Wavegraph project),
and by CNRS PICS/Inde. VG acknowledges Inspire division, DST,
Government of India for the fellowship support. VG would like to thank
IISER-TVM for providing facility to complete the initial part of this
work.  VG thanks to Max Planck Partner Group on Gravitational Waves
grant for the travel support to Albert Einstein Institute. The
authors are grateful to the LIGO Scientific Collaboration and Virgo
Collaboration for giving access to the simulation software used here,
and specifically to the cWB team for their help and support. The
authors are grateful to Sergey Klimenko, Marco Drago, St\'ephane
Jaffard, and Aur\'elia Fraysse for their valuable comments and
suggestions.

\bibliography{reference}

\end{document}